\documentclass[11pt,a4paper]{article}
\pdfoutput=1
\usepackage{jheppub}
\usepackage{graphicx}
\usepackage{epsfig}
\usepackage{wrapfig}
\usepackage{amsmath}
\usepackage{amssymb}
\usepackage{dsfont}

\newcommand{\expv}[1]{\left \langle #1 \right \rangle}

\newcommand{\Wuppertal}{Institute for Theoretical Physics, Bergische Universit\"at Wuppertal, Gau{\ss}str. 20, D-42119, Germany.}
\newcommand{\Regensburg}{Institute for Theoretical Physics, Universit\"at Regensburg
Universit\"atsstr 31, D-93040 Regensburg, Germany.}
\newcommand{\Budapest}{Institute for Theoretical Physics, E\"otv\"os University
P\'azm\'any P. 1, H-1117 Budapest, Hungary.}
\newcommand{\Juelich}{J\"ulich Supercomputing Center, Forschungszentrum J\"ulich
Wilhelm-Johnen-Str. D-52425 J\"ulich, Germany.}

\newcommand{\Lambdamsbar}{\Lambda_{\overline{\rm MS}}}
\newcommand{\be}{\begin{equation}}
\newcommand{\ee}{\end{equation}}
\newcommand{\Tr}{\textmd{Tr}}
\renewcommand{\d}{\textmd{d}}

 \preprint{WUB/12-09}
\title{
Precision SU(3) lattice thermodynamics for a large temperature range}
\author[1]{Sz.~Bors\'anyi,}
\author[2]{G.~Endr\H{o}di,}
\author[1,3,4]{Z.~Fodor,}
\author[3]{S.~D.~Katz,}
\author[1]{K.~K.~Szab\'o}
\affiliation[1]{\Wuppertal}
\affiliation[2]{\Regensburg}
\affiliation[3]{\Budapest}
\affiliation[4]{\Juelich}

\emailAdd{borsanyi@uni-wuppertal.de}
\emailAdd{gergely.endrodi@physik.uni-regensburg.de}
\emailAdd{fodor@physik.uni-wuppertal.de}
\emailAdd{katz@bodri.elte.hu}
\emailAdd{szaboka@general.elte.hu}

\abstract {
We present the equation of state (pressure, trace anomaly, energy density and entropy density) of the $\mathrm{SU}(3)$
gauge theory from lattice field theory in an unprecedented precision and 
temperature range. We control both finite size and cut-off effects.
The studied temperature window ($0.7\dots 1000\, T_c$) stretches from the
glueball dominated system into the perturbative regime, which allows us
to discuss the range of validity of these approaches. We also determine
the preferred renormalization scale of the Hard Thermal Loop scheme and
we fit the unknown $g^6$ order perturbative coefficient at extreme high
temperatures $T>100\,T_c$.  We furthermore quantify the nonperturbative contribution to the trace anomaly using 
a simple functional form.
Our high precision data allows one to have a complete theoretical description of the equation of state from $T=0$ all the way to the phase
transition, through the transition region into the perturbative regime up to the Stefan-Boltzmann limit. We will discuss this description, too.
}

\arxivnumber{1204.6184}

\begin{document}

\maketitle

\section{Introduction}

The general feature of asymptotic freedom makes weak coupling approaches
very natural in non-abelian gauge theories, such as the $\mathrm{SU}(3)$ model, which
describes the gluonic degrees of freedom of Quantum Chromodynamics. At
asymptotically high temperatures low orders of perturbation theory may 
be acceptable, but at any lower scale that could be probed by a realistic
experiment an extension is necessary: either by the inclusion of very high order
diagrams, or by an efficient resummation scheme, such as Hard Thermal Loops (HTL).
Note however that analytic perturbative expansions are plagued by
infrared divergences due to which the series can be computed only up
to a given finite order.
There is strong simulation evidence that at low temperatures
($T_c\sim 260$ MeV) the gluonic matter freezes and a first order transition takes
place. At even lower temperatures colorless non-perturbative excitations govern
the thermodynamics. To describe the phase transition or the glueball gas no
weak coupling scheme succeeds and one has to rely on a natively non-perturbative
approach, such as lattice field theory.
The pure gauge theory is a very good test-bed for perturbative and non-perturbative studies. On the one hand it is only moderately CPU demanding. Full QCD for $\mu=0$ is more expensive, and at $\mu>0$ it is even more expensive (see e.g.~\cite{Fodor:2001au}). On the other hand the pure $\mathrm{SU}(3)$ theory shows all the infrared difficulties of the full theory.

The past years witnessed considerable achievements on the side of the analytical
results (see e.g. \cite{Blaizot:2003iq}). HTL perturbation theory (which was first developed in Refs.~\cite{Braaten:1989kk,Andersen:1999fw}) has been recently used to calculate the pure
$\mathrm{SU}(3)$ gauge theory's thermodynamic potential to the next-to-next-to-leading
order (NNLO)~\cite{Andersen:2009tc,Andersen:2010ct}. The authors used their
results at intermediate temperatures ($\sim 4\,T_c$) where existing lattice data
was available. Later the same authors have extended their results to full QCD
(with massless quarks)~\cite{Andersen:2010wu,Strickland:2010tm}, which was
compared to results of the Wuppertal-Budapest collaboration
\cite{Borsanyi:2010cj}.

In the weak coupling expansion even higher orders can be computed by dimensional reduction~\cite{Braaten:1995cm}. This method was applied to calculate the pressure of QCD first in Refs.~\cite{Braaten:1995ju,Braaten:1995jr} up to $g^5$.
When re-expanded in the coupling $g$ the full expression can be calculated up to $g^6\log(g)$ order and was given in Ref.~\cite{Kajantie:2002wa}
%In conventional perturbation theory even higher orders can be computed by
%dimensional reduction. The full expression up to  $g^6\log(g)$ order is given in~\cite{Kajantie:2002wa}
and compared to the Bielefeld lattice data \cite{Boyd:1996bx} at $T=4.5\,T_c$. Fitting the
pressure (thermodynamic potential) the slope of the pressure curve was successfully
predicted. This raised hope that at this high order perturbation theory
does possess some predictive power at phenomenological temperatures. In this paper
we repeat this fitting procedure at a much higher temperature, where the sixth
order can be shown to be a minor correction to the fifth order.

For more than a decade the renowned paper by Boyd et al~\cite{Boyd:1996bx} has
been the reference lattice simulation of the $\mathrm{SU}(3)$ theory in the temperature
range of $1\dots4.5 \,T_c$. It uses the plaquette gauge action at up to $N_t=8$
lattice spacing and an aspect ratio of 4. Here $N_t$ denotes the number of
lattice points in the Euclidean time direction, meaning that the lattice spacing
at any given temperature $T$ is $a=1/(TN_t)$. The fixed $N_t$ approach has been
introduced in Ref.~\cite{Engels:1990vr} and this work follows it, too. It
implies that the lattice spacing varies with temperature.
Continuum limit is achieved by performing an $1/N_t\to0$ extrapolation on the
data at a set of fixed physical temperatures.  The aspect ratio $r=LT$ sets the
ratio between space and time-like lattice points.

Since the publication of~\cite{Boyd:1996bx} several similar simulations were performed to study pure gauge theory. The equation of state has been recalculated using the Symanzik improved
gauge action~\cite{Beinlich:1997ia}.
This set of simulations have been further generalized
to SU($N_c$) theories with $N_c>3$ in Refs.~\cite{Panero:2009tv,Datta:2009jn}.
Alternatively, the equation of state can also be calculated by fixing the lattice
spacing, and using $N_t$ for tuning the temperature~\cite{Umeda:2008bd}. This
approach is mostly advantageous with Wilson-type dynamical fermions, and less
economic for the pure gluonic theory.

In most fixed $N_t$ simulation projects, like Ref.~\cite{Boyd:1996bx}, the aspect ratio is kept constant to allow the use of a single lattice geometry. This means that higher temperatures are simulated at smaller volumes. As we discuss later the aspect ratio sets the maximum temperature as a precondition for the non-perturbativeness of the simulation: in first approximation one expects $T\lesssim r T_c$.  In most previous works this was set to $r=4$.

The outline of the paper can be summarized as follows. In Sec.~\ref{sec:EoSdef} we
briefly present the lattice framework for the equation of state. Secs.~\ref{sec:nonpert},~\ref{sec:finitevol} and~\ref{sec:scaleset} deal with some technicalities, namely with the determination of the non-perturbative/non-ideal contributions, with finite volume effects and with scale setting. Readers who are not interested in these details
can jump to Sec.~\ref{sec:results} for our results or even to Sec.~\ref{sec:models} for a short
summary of our findings and for comments on the various theoretical descriptions and model building.

\section{Equation of state}
\label{sec:EoSdef}

In the lattice simulation we use the Symanzik improved gauge action~\cite{Curci:1983an,Luscher:1985zq},
\be
S_g=-\beta \left[ c_0 \sum\limits_{n,\mu<\nu} \textmd{Re} \, \Tr \, U^{1\times1}_{\mu \nu}(n)  + c_1 \sum\limits_{n,\mu\ne \nu} \textmd{Re}\,\Tr \, U^{2\times1}_{\mu \nu}(n) \right]
\label{eq:symanzikact}
\ee
with inverse gauge coupling $\beta=6/g^2$. Here the coefficients are set to $c_0=5/3$ and $c_1=-1/12$, such that the scaling with the lattice spacing $a$ is improved on the tree level.
The primary observable in our approach is the trace anomaly\footnote{The trace anomaly is often also called interaction measure as it measures the deviation from the equation of state of an ideal gas $\epsilon=3p$.}, as measured on the lattice~\cite{Engels:1990vr},
\be
\frac{I}{T^4}\equiv \frac{\epsilon-3p}{T^4}= N_t^4 a\frac{\d \beta}{\d a} \left( \expv{s_g}_{N_s^3\times N_t} - \expv{s_g}_{N_s^3\times N_t^{\rm sub}} \right)
\label{eq:defI}
\ee
where $\beta(a)$ is the relation between the bare coupling and the lattice spacing $a$ which we determine in section~\ref{sec:scaleset}. 
The gauge action density $s_g=T/V\cdot S_g$ contains quartic divergences in the cutoff, which we cancel here by taking the difference between measurements at the same parameter $\beta$ but different temporal extent $N_t$, i.e. different temperature. The temporal size $N_t^{\rm sub}$ of the lattice used for the subtraction will be either set to $2\cdot N_t$ (corresponding to half the temperature) or $N_s$ (corresponding to zero temperature), as will be discussed later.

Having calculated the trace anomaly as a function of the temperature, all other thermodynamic observables can also be reconstructed. The pressure is obtained as a definite integral,
\be
\frac{p(T)}{T^4}-\frac{p(T_0)}{T_0^4} = \int\limits_{T_0}^T {\frac{I(T')}{T'^5} \d T'},
\label{eq:pfromI}
\ee
where the integration constant can be set using a glueball resonance gas model, see section~\ref{sec:glueballgas}.
Using the pressure and the trace anomaly, the energy density $\epsilon$ and the entropy density $s$ can be calculated as
\be
\epsilon= I + 3p,\quad\quad s = \frac{\epsilon+p}{T}.
\label{eq:eosq}
\ee

Besides the thermodynamic observables defined above, for the setting of the lattice scale we also measure the susceptibility $\chi_P$ of the Polyakov loop $P$, defined as
\be
P = \frac{1}{V} \sum\limits_{n_1,n_2,n_3} \Tr \prod\limits_{n_4=0}^{N_t-1} U_4(n), \quad\quad
\chi_P = V \left( \expv{P^2} - \expv{P}^2 \right).
\label{eq:Polyakov}
\ee

\section{Non-perturbative contributions}
\label{sec:nonpert}

\begin{wrapfigure}{r}{8.4cm}
\centering
\vspace*{-0.4cm}
\includegraphics*[width=7.8cm]{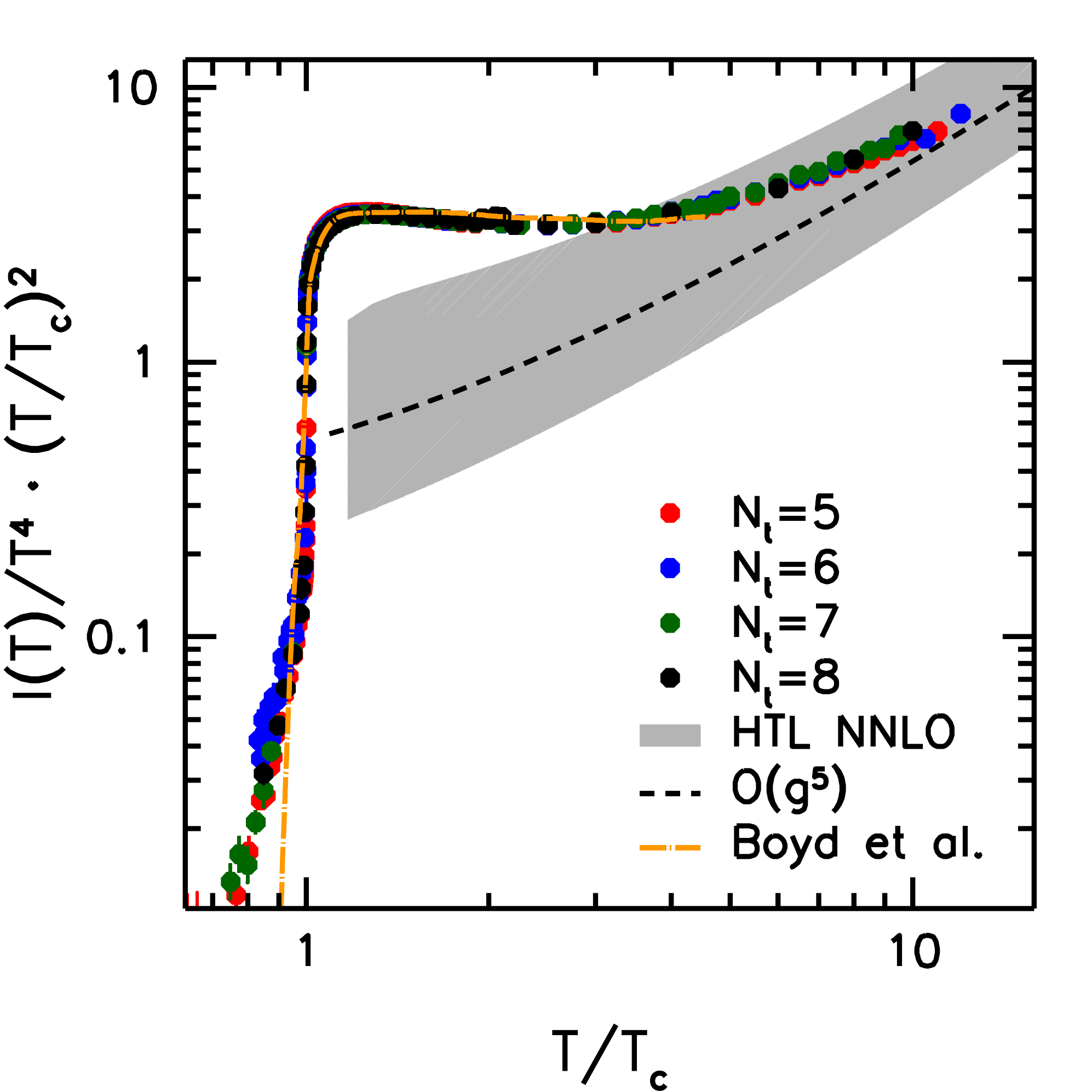}
\vspace*{-0.1cm}
\caption{Our results for the normalized trace anomaly multiplied by $T^2/T_c^2$ for $N_t=5,6,7$ and $8$ (red, green and blue dots, respectively). Also plotted are lattice results of~\cite{Boyd:1996bx}, $g^5$ perturbation theory~\cite{Kajantie:2002wa} and the HTL approach~\cite{Andersen:2010ct}.}
\label{fig:tracea_NP}
\vspace*{-0.3cm}
\end{wrapfigure}

Recently there have been interesting observations about the presence of a non-perturbative contribution in the equation of state in the transition region~\cite{Meisinger:2001cq,Pisarski:2006yk}. For dimensional reasons, any finite order perturbative formula can only give logarithmic corrections to the $p(T)\sim T^4$ Stefan-Boltzmann law.
Instead of such logarithmic corrections, lattice data suggests that there is an approximately quadratic contribution which is dominant for temperatures up to $\sim4\,T_c$. 
This non-perturbative pattern may be explained within a fuzzy bag model~\cite{Pisarski:2006yk}, in terms of a dimension-2 gluon condensate~\cite{Pisarski:2000eq,Kondo:2001nq}, in a system of transversely polarized quasi-particles~\cite{Castorina:2011ja} or within the gauge/string duality~\cite{Andreev:2007zv}. Here we do not go into the viability of such models and only identify it as the dominant non-perturbative contribution.

This non-perturbative contribution can be best observed in the trace anomaly $I=\epsilon-3p$. Specifically, it is instructive to study the combination $I/T^4\cdot(T/T_c)^2$, which is shown in Fig.~\ref{fig:tracea_NP}. Our results with the Symanzik improved gauge action for various lattice spacings are compared to results obtained with the Wilson gauge action~\cite{Boyd:1996bx}, the 4-loop perturbative expansion~\cite{Kajantie:2002wa} and the HTL NNLO scheme. While for the former the renormalization scale $\mu=2\pi T$ is used (black dashed line in the figure), for the latter a range of $\mu_{\rm HTL}=\pi T \ldots 4\pi T$ is considered (gray band).

Apparently, the combination $I/T^2$ as measured on the lattice is approximately constant in the temperature range $T_c<T<5\,T_c$ (there are however discrepancies between the Symanzik and Wilson results, see discussion later). While up to $5\,T_c$ lattice results seem completely incompatible with the perturbative predictions, at larger temperatures our results also account for the $T^4$-like steep rise in $I(T)$ indicating a qualitative agreement with perturbative methods. This suggests that besides the ideal (perturbative) contribution $\sim T^4$, $I$ also contains a non-ideal (non-perturbative) term $\sim T^2$. Thus we separate the trace anomaly into two parts,
\be
\frac{I(T)}{T^4} = \frac{I_{\rm pert}(T)}{T^4} + \frac{I_{\rm np}(T)}{T^4}.
\ee
The pressure can be obtained from the trace anomaly with a definite integral, as in~(\ref{eq:pfromI}). At extremely high temperatures its value is given by the Stefan-Boltzmann limit $p_{\rm SB} = 8\pi^2/45\,T^4$. Integrating down from this point one obtains,
\be
\frac{p(T)}{T^4} = p_{\rm SB} - \int\limits_{T}^\infty \left(\frac{I_{\rm pert}(T')}{T'^5} + \frac{I_{\rm np}(T')}{T'^5} \right) \d T' = \frac{p_{\rm pert}(T)}{T^4} - \int\limits_T^\infty \frac{I_{\rm np}(T')}{T'^5} \d T'.
\label{eq:pressureformula}
\ee

The results for the trace anomaly in the high-temperature region allow for a fitting of the HTL renormalization scale $\mu_{\rm HTL}$ and the unknown coefficient ($q_c$ in the notation of~\cite{Kajantie:2002wa}) of the $\mathcal{O}(g^6)$ order contribution of perturbation theory. While $q_c$ has already been calculated by means of a fit to the lattice data of~\cite{Boyd:1996bx}, here we are able to repeat this fitting procedure at a much higher temperature, where the sixth order can be shown to be a minor correction to the fifth order.
Once the optimal coefficient of the $g^6$ term is known, the non-perturbative contribution can also be quantified through a fit to some specified parameters of the function $I_{\rm np}(T)$.

\section{Finite volume effects and high temperature}
\label{sec:finitevol}

Existing lattice results for the pressure end at around $5\,T_c$. These include results in the pure gauge sector with the Wilson plaquette action~\cite{Boyd:1995zg,Boyd:1996bx} and also with various improved actions like the Symanzik action~\cite{Beinlich:1995ik,Beinlich:1997ia}, renormalization group-improved actions~\cite{Okamoto:1999hi} or fixed-point actions~\cite{Papa:1996an}.
The effect of changing the number of colors~\cite{Panero:2009tv} was also studied.
Results for the pressure of full QCD~\cite{Karsch:2000ps, Bernard:2006nj, Aoki:2005vt,Cheng:2009zi,Borsanyi:2010cj} are also present only up to $(5-10)\cdot T_c$.

There are two main reasons for the absence of high temperature results: first, at increasingly high temperatures the signal/noise ratio in the trace anomaly decreases significantly, dropping below $0.01\%$ already above $T_c$. Consequently, it becomes more and more difficult to detect a nonvanishing value for $I/T^4$, and this small signal is just the information necessary to fit the unknown perturbative parameters like $q_c$ mentioned in the previous section. Second, since the lattice spacing varies with the temperature as $a=(N_t T)^{-1}$, in order to have a constant physical lattice size, the number of lattice points $N_s$ in the spatial directions in principle has to increase like $T$. While the former problem can be avoided by accumulating larger statistics, the latter obstacle is more of a matter of principle. 
Length scales discussed in the HTL approach and in the dimensional reduction method are normally well accommodated in the lattice.
However, to establish the range of validity of the
perturbative approach itself, one has to simulate the non-perturbative
$\sim T_c$ scale, too. This implies that the aspect ratio $N_s/N_t$ has to be increased linearly with $T$, up to temperatures where the matching to perturbation theory can be performed in a reliable manner.

Keeping in mind these considerations we perform three sets of simulations. First, we calculate the trace anomaly in the temperature range of $T/T_c= 0.7 \ldots 15$ (on $80^3 \times 5$, $96^3 \times 6$ and $112^3 \times 7$ lattices) and also extract a continuum limit from these results. These lattices with aspect ratio $N_s/N_t=16$ accommodate the non-perturbative scale $T_c^{-1}$ up to approximately $16\,T_c$.
We also support the continuum extrapolation with an additional $N_t=8$ set of lattices ($64^3\times 8$) below $8\,T_c$. This combined extrapolation is described in the beginning of section~\ref{sec:results}. From the trace anomaly various other thermodynamic functions can be determined according to the thermodynamic relations~(\ref{eq:pfromI}) and~(\ref{eq:eosq}).

As a next step we study the finite volume scaling of the trace anomaly on a non-continuum data set at $N_t=5$. We presents results using lattices of aspect ratio $N_s/N_t=4, 6, 8, 16$ and $24$. The latter $120^3 \times 5$ lattice accommodates the $T_c^{-1}$ scale up to $24\,T_c$. Using these results we test finite size effects in the whole temperature region, see subsection~\ref{sec:voldep_tracea}. 

Since from this analysis we find that finite size effects are smaller than our statistical errors provided that $N_s/N_t\ge6$, in our third set of simulations we calculate the continuum equation of state in a somewhat smaller box ($r=8$ on $40^3 \times 5$, $48^3 \times 6$ and $64^3 \times 8$ lattices) up to $1000\,T_c$. In subsections~\ref{sec:fitg6} and~\ref{sec:fithtl} we use this data set to find the optimal free parameters of existing perturbative calculations, i.e. the already mentioned $q_c$ parameter of $\mathcal{O}(g^6)$ perturbation theory and the renormalization scale $\mu_{\rm HTL}$ of the HTL scheme. Using these small volume results we observe a good agreement with the newly fitted perturbative formulae, indicating that this approach successfully connects the low temperature non-perturbative region with the high temperature perturbative realm. The precision of our data points exceeds any previous calculation by about an order of magnitude.

In order for the large lattices to fit in the memory of our computer system, the renormalization of the trace anomaly was done via half-temperature subtraction, as explained in section~\ref{sec:EoSdef}. Specifically, we calculate
\be
\frac{I(T)}{T^4} = \left(\frac{I(T)}{T^4} -\frac{1}{16}\frac{I(T/2)}{(T/2)^4}\right) + \frac{1}{16}\frac{I(T/2)-I(0)}{(T/2)^4}.
\label{eq:traceameas}
\ee
In the right hand side of the expression $I(T)$, $I(T/2)$, $I(T/2)$ and $I(0)$ are obtained by using lattice sizes of $(2N_s)^3\times N_t$, $(2N_s)^3\times (2N_t)$ at a given $\beta$ and $N_s^3\times N_t$ and $N_s^4$ at another $\beta'$, respectively. The lattice spacing at $\beta$ is half the lattice spacing at $\beta'$. This guarantees that the spatial volumes are the same. The transition in the pure $\mathrm{SU}(3)$ theory is a first order~\cite{Fukugita:1989yb} phase transition (in contrast to full QCD which has a crossover for physical masses~\cite{Aoki:2006we}). Due to this first order phase transition the trace anomaly depends on the volume around the critical temperature. In order to account for this dependence we used the above choice for the volumes. Thus both terms are measured with the same physical volume which ensures that the sum is smooth around $T=2\,T_c,4\,T_c,\ldots$ (otherwise the difference between the volumes shows up as small cusps at these temperatures). For the lattices with half the spatial size (i.e. the second term in the right hand side of~(\ref{eq:traceameas})) the subtraction is carried out in the standard way, i.e. at $T=0$. The continuum limit from this combined technique is equal to what one finds using the standard scheme.

\section{Scale setting}
\label{sec:scaleset}

Besides the proper treatment of finite volume effects another challenging issue was the accurate determination of the non-perturbative beta function corresponding to the Symanzik improved action. 
This function determines the relation between the lattice spacing and
the bare gauge coupling which is the only free parameter in the lattice Lagrangian.
The standard strategy for obtaining the lattice scale is the determination of the string tension $\sigma$ or the Sommer parameter $r_0$ in a zero temperature setting. Yet, for the fine lattices we needed for the high temperatures this would have been computationally extremely demanding.
Instead it was advantageous to define the lattice spacing in terms of the transition temperature. To this end we determined the critical couplings
$\beta_c$ up to $N_t = 20$ from the peak of the Polyakov loop susceptibility~(\ref{eq:Polyakov}).
For finer lattices we determined the scale using the continuum extrapolated value of the renormalized Polyakov loop at $T = 1.5\,T_c$, analogously to the step scaling method. This allowed us to calculate the critical coupling up to $N_t = 36$.

Matching to the universal two-loop running (in terms of the improved
coupling in the ``E'' scheme \cite{Bali:1992ru}, generalized for the case of
the Symanzik improved action) we determined the lambda
parameter in terms of the transition temperature:  $T_c/\Lambdamsbar=1.26(7)$.
(The error
is overwhelmingly systematic and reflects the sensitivity to various continuum
extrapolations.) This is consistent with the combination of previous determinations:
the Lambda parameter $\Lambdamsbar=0.614(2)(5) r_0^{-1}$  of~\cite{Gockeler:2005rv}
can be translated to $\sqrt{\sigma}$ units using $\sqrt{\sigma}r_0=1.192(10)$
(based on~\cite{Guagnelli:1998ud}) and then used with
$T_c/\sqrt{\sigma}=0.629(3)$ of~\cite{Boyd:1996bx}. Through our direct
result one can easily translate the scale setting of the perturbative
expressions to the lattice language.

\section{Results}
\label{sec:results}

\begin{figure}[h!]
\centering
\vspace*{-0.2cm}
\includegraphics*[width=12cm]{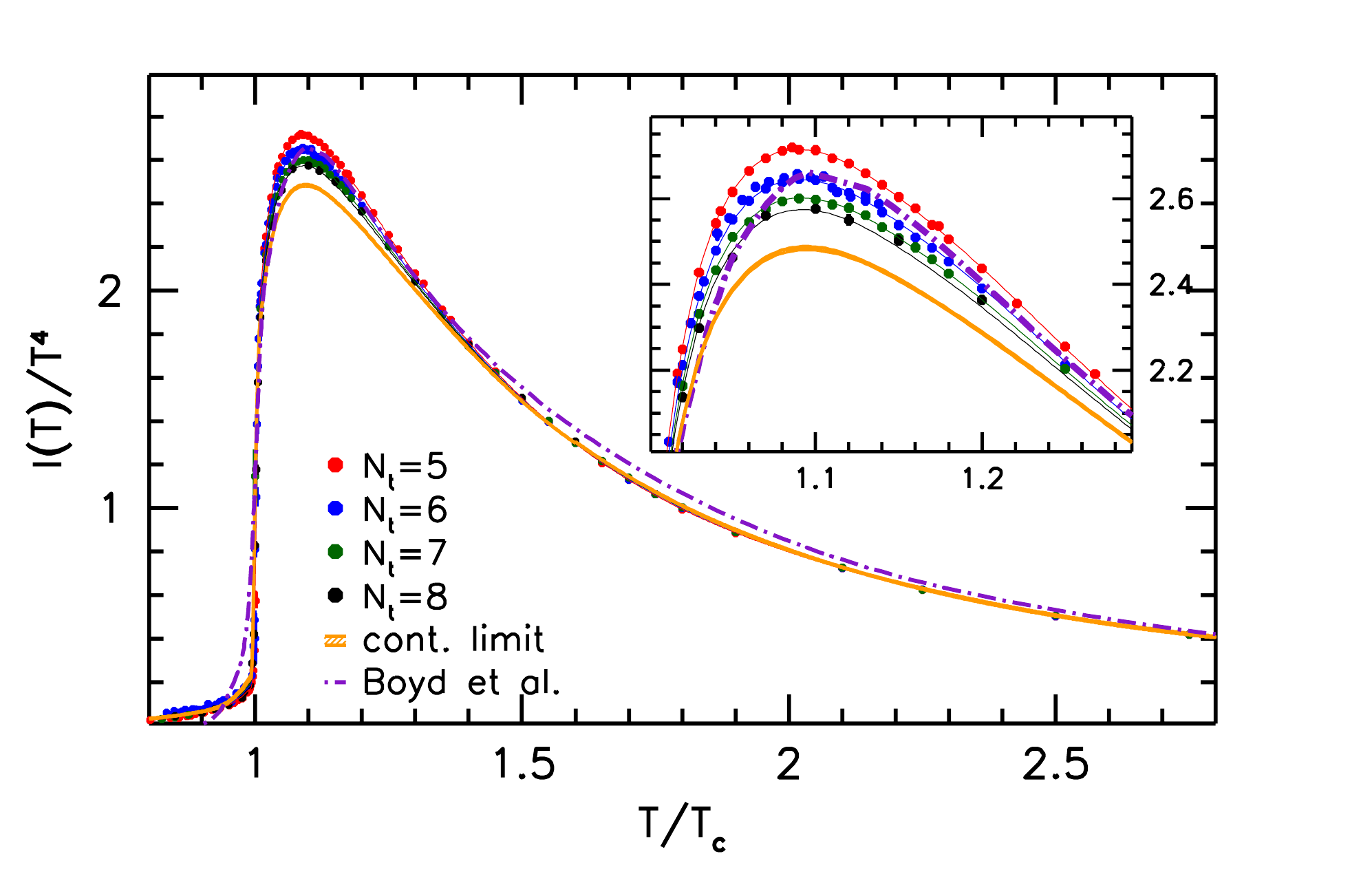}
\vspace*{-0.1cm}
\caption{
The trace anomaly on $N_s/N_t=8$ lattices for various lattice spacings in the transition region. The result of a combined spline fit for each lattice spacing, together with the continuum extrapolation is shown by the colored lines and the yellow band, respectively. For comparison results with the standard Wilson action~\cite{Boyd:1996bx} are also shown by the dashed-dotted line.
The continuum estimate of~\cite{Boyd:1996bx} in the inset has the same peak height as our $N_t=6$ curve, which is about $7\%$ higher
than our continuum value.}
\vspace*{-0.3cm}
\label{fig:i_trans}
\end{figure}

First we reproduce the results of~\cite{Boyd:1996bx} in the transition region. In Fig.~\ref{fig:i_trans} these results are compared to the trace anomaly measured on our first set of simulations, i.e. on large lattices ($N_s/N_t=16$) with $N_t=5,6,7$, supplemented by $N_t=8$, with $N_s/N_t=8$. From these four sets of results we perform a continuum extrapolation via a combined spline fitting method. The datasets for different lattice spacings are fitted together by an $N_t$-dependent spline function. This ``multi-spline'' function -- defined upon a set of nodepoints $\beta_k$ with $k=1\ldots K$ -- is parameterized by two values at each nodepoint, written in the form $a_k+b_k N_t^{-2}$ (the $N_t$-dependence is motivated by the scaling properties of the Symanzik action). We fit these $2K$ parameters to the measurements: the minimum condition for $\chi^2$ leads to a set of linear equations, which can be solved for the parameters.

As a result we have a smooth function interpolating our data for each $N_t$ (colored lines in the figure), together with a smooth, continuum extrapolated curve (yellow band in the figure). 
The statistical error of the fit is determined by a jackknife analysis, while the systematic error of the continuum result is estimated by the difference between the extrapolation from $N_t=5,6,7$ and $N_t=5,6,7,8$.
As visible in the figure, data points for various lattice spacings are on top of each other, with the exception of the transition region. This region is zoomed into in the inset of the figure, showing that our data indeed exhibits the expected scaling.

As Fig.~\ref{fig:i_trans} shows there is an apparent discrepancy between our continuum result and that of~\cite{Boyd:1996bx}, particularly around $T_c$. 
In Ref.~\cite{Boyd:1996bx} the Wilson gauge action was applied and the continuum limit of the trace anomaly was calculated based on $N_t=6$ and $8$ lattices. Fig.~\ref{fig:i_trans} shows that even with the Symanzik improved action there is significant difference between the $N_t=8$ data (black points) and the continuum curve (yellow band) for temperatures just above $T_c$. The extrapolation using several lattice spacings is therefore essential in this temperature region. Moreover, differences can also be attributed to finite volume effects as well as to the systematics of the scale setting procedures.

\subsection{Comparison to the glueball gas model}
\label{sec:glueballgas}

\begin{figure}[ht!]
\centering
\vspace*{-0.3cm}
\includegraphics*[width=12cm]{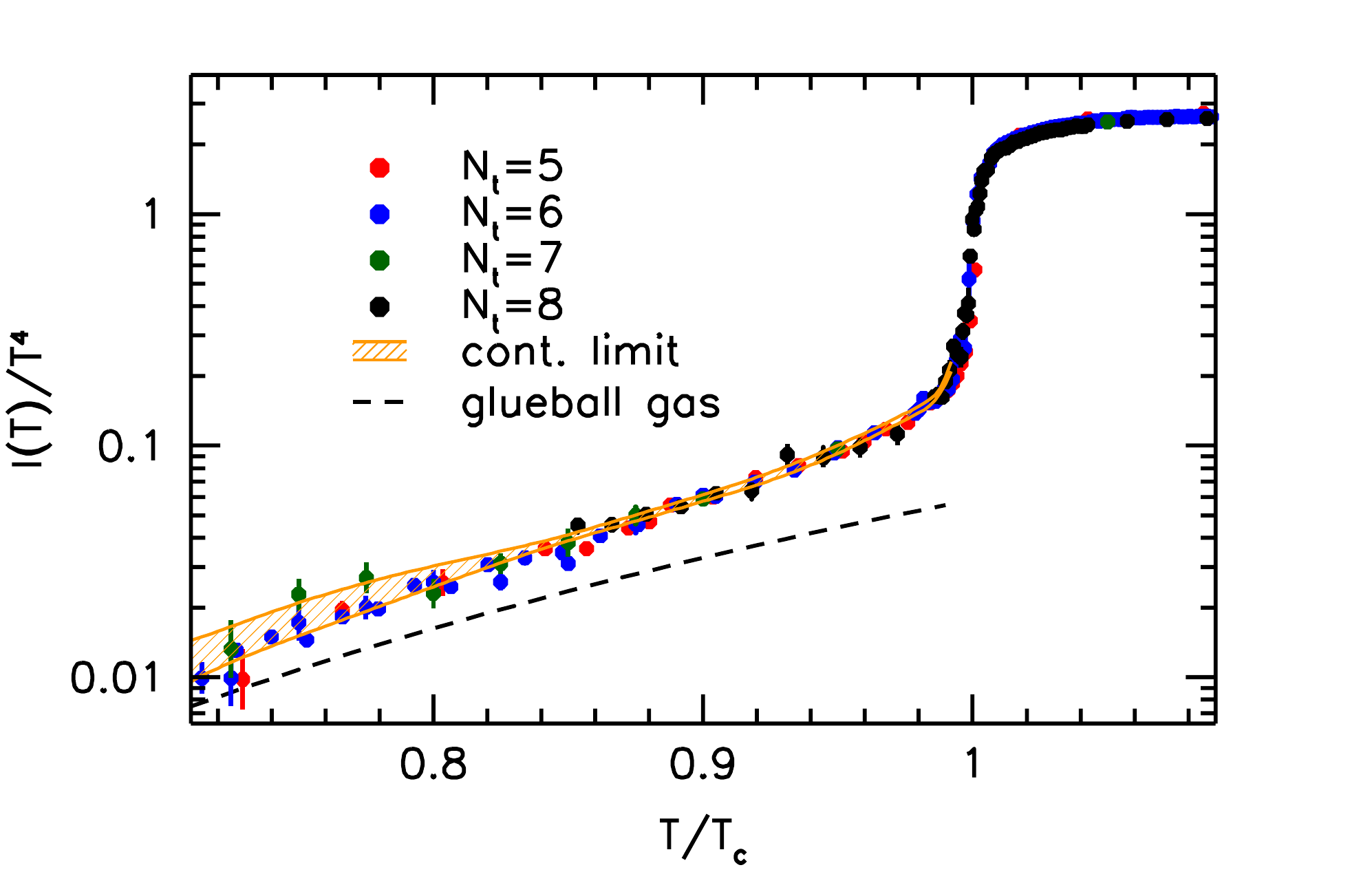}
\vspace*{-0.3cm}
\caption{The trace anomaly in the confined phase measured with various lattice spacings and the continuum extrapolation (yellow band). The dashed line corresponds to the glueball resonance model, estimated from the twelve lightest glueballs.
}
\label{fig:coldtra}
\vspace*{-0.1cm}
\end{figure}

\begin{figure}[ht!]
\center
\includegraphics[width=12cm]{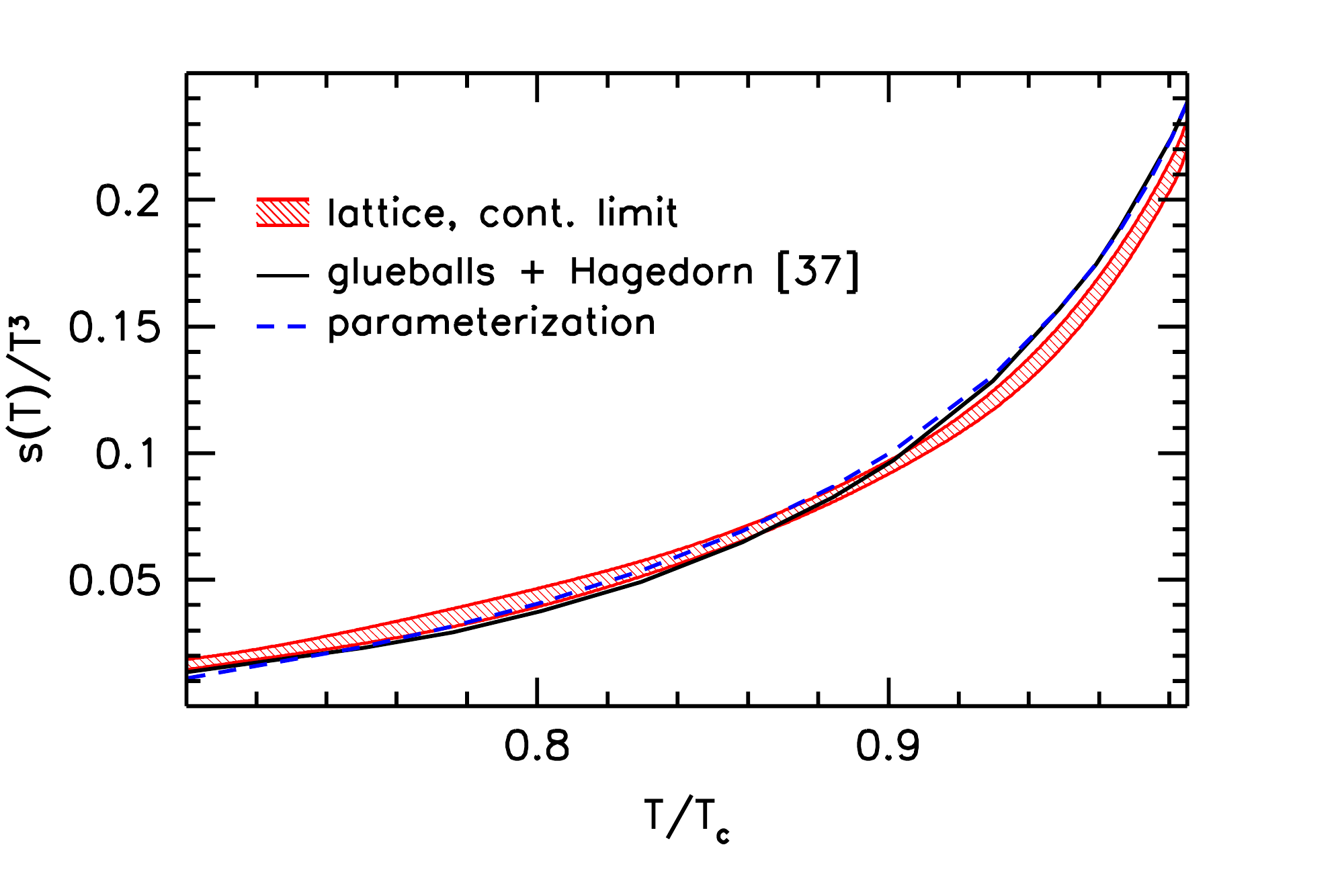}
\vspace*{-.5cm}
\caption[]{
Entropy in the confining phase. The red band shows our continuum extrapolated
lattice result based on $N_t=5,6$ and $8$ data. The thick line is the entropy
of a glueball gas where the Hagedorn spectrum is assumed beyond the
two-particle threshold~\cite{Meyer:2009tq}. The dashed line shows our
parameterization in Eq.~(\ref{eq:scoldparam}).
}
\label{fig:hagedorn}
\end{figure}

In order to explore the thermodynamics of the confined phase, next we zoom into the low temperature region $T<T_c$ in the left panel of Fig.~\ref{fig:coldtra}. In this region one can also calculate the trace anomaly within the glueball resonance model (note that the hadron resonance gas model works very
well for full QCD~\cite{Borsanyi:2010bp,Borsanyi:2010cj}). In Fig.~\ref{fig:coldtra} we plot our results together with the contribution of the first twelve glueballs of~\cite{Chen:2005mg}. There is an apparent deficit of the model prediction as compared to the lattice results. 
It has been suggested~\cite{Meyer:2009tq} to cover this deficit with the addition of a  Hagedorn spectrum~\cite{Hagedorn:1965st} contribution $\rho(M)\propto \exp(M/T_h)$. As it can be seen in Fig.~\ref{fig:hagedorn} the temperature dependence of our continuum extrapolated equation of state shows a good agreement with this picture (we actually use the direct lattice data of~\cite{Meyer:2009tq} to set the integration constant of the entropy density). We parameterized the result of this theoretical description (glueballs + Hagedorn spectrum) and comment on it in the last section of our paper. 
	
\subsection{Volume dependence of the results}
\label{sec:voldep_tracea}

\begin{figure}[h!]
\centering
\vspace*{-0.2cm}
\includegraphics*[width=12cm]{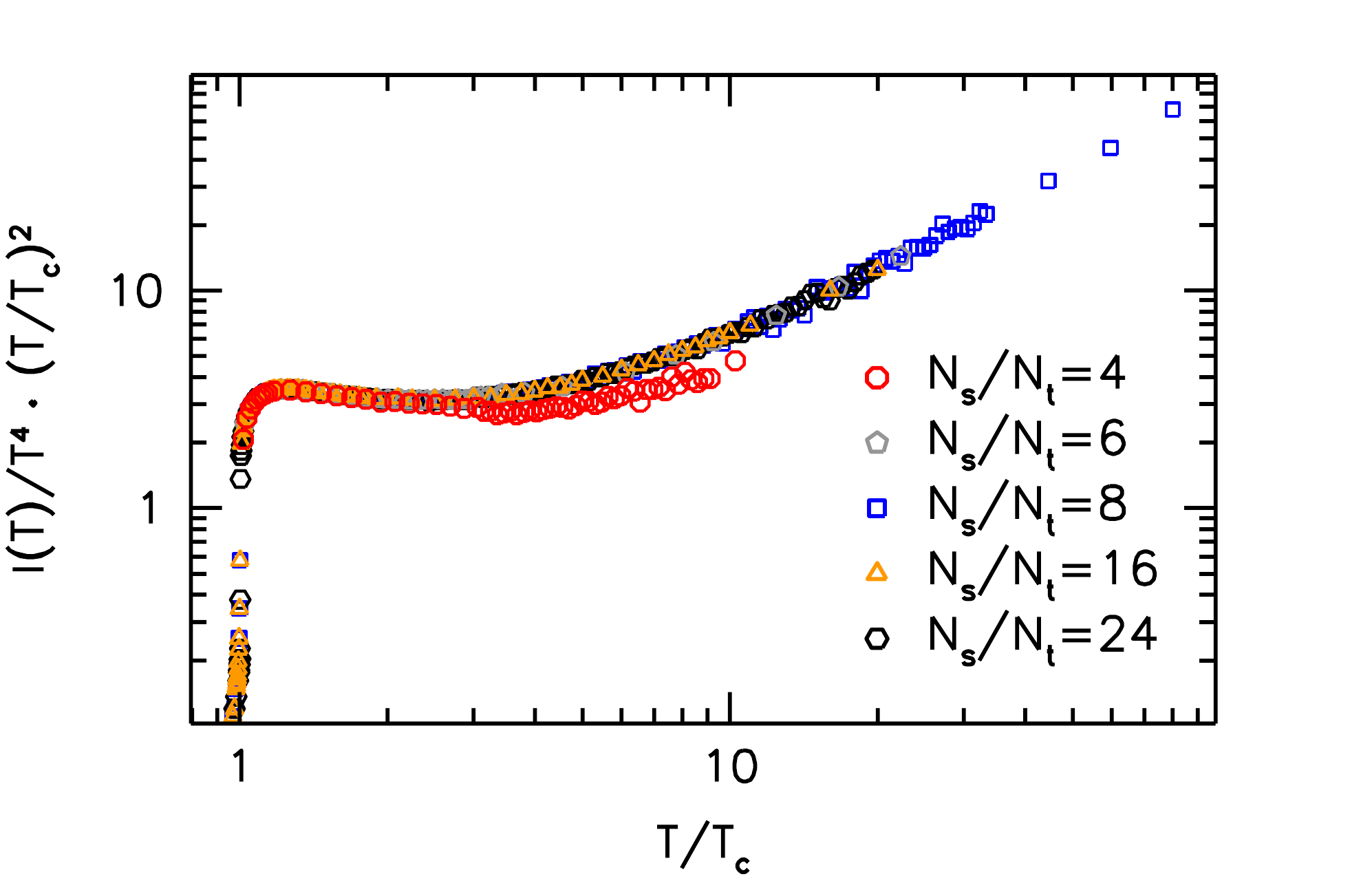}
\vspace*{-0.cm}
\caption{Volume dependence of the trace anomaly on our $N_t=5$ lattices. Unless the box is very small, there is no significant
difference whether or not the box size allows contributions from the inverse $T_c$ scale.}
\label{fig:voldep}
\vspace*{-0.0cm}
\end{figure}

As discussed in section~\ref{sec:nonpert}, the trace anomaly contains a non-perturbative contribution which dominates for $T_c<T<5 \,T_c$. 
The effect of this $\sim T^2$ contribution reduces at increasing temperatures. Moreover, the presence of this contribution becomes unnoticeable at sufficiently high $T$, regardless of whether or not the lattice size accommodates the inverse $T_c$ scale. One way to discuss the relevance of this non-perturbative scale is to compare the trace anomaly at various spatial volumes. This comparison is shown in Fig.~\ref{fig:voldep} for our $N_t=5$ lattices. 
The standard aspect ratio $N_s/N_t=4$ gives somewhat smaller values for $I/T^4$, but beyond $N_s/N_t=6$ we do not see any difference in the results above the transition region.

We summarize our findings as i) the large volume lattice trace anomaly data shows qualitative (and as 
we find using the fitted $g^6$ order coefficient, also quantitative, see later) agreement with the perturbative
results for $T>10\,T_c$, and ii) we see no deviation between results
from various volumes (with $N_s/N_t\ge6$), moreover iii) the dominant non-perturbative
contribution loses significance as $\sim 1/T^2$.
These considerations suggest that -- even if the lattice volumes are ever shrinking as the temperature is increased -- our results are able to describe the physical trace anomaly 
(and its integral, the thermodynamic potential) within the error bars shown.
Of course, this assumes that all relevant scales are properly accounted for.  
In all our lattices besides the hard ($T$) scale, the soft ($gT$) as well as ultrasoft ($g^2T$) scales are well represented. Thus it is reasonable to conjecture that our $N_s/N_t=8$ dataset reliably connects the transition region with the perturbative regime.

\subsection{Fitting improved perturbation theory}
\label{sec:fitg6}

\begin{figure}[h!]
\centering
\vspace*{-0.2cm}
\mbox{
\hspace*{-0.3cm}
\includegraphics[width=8.2cm]{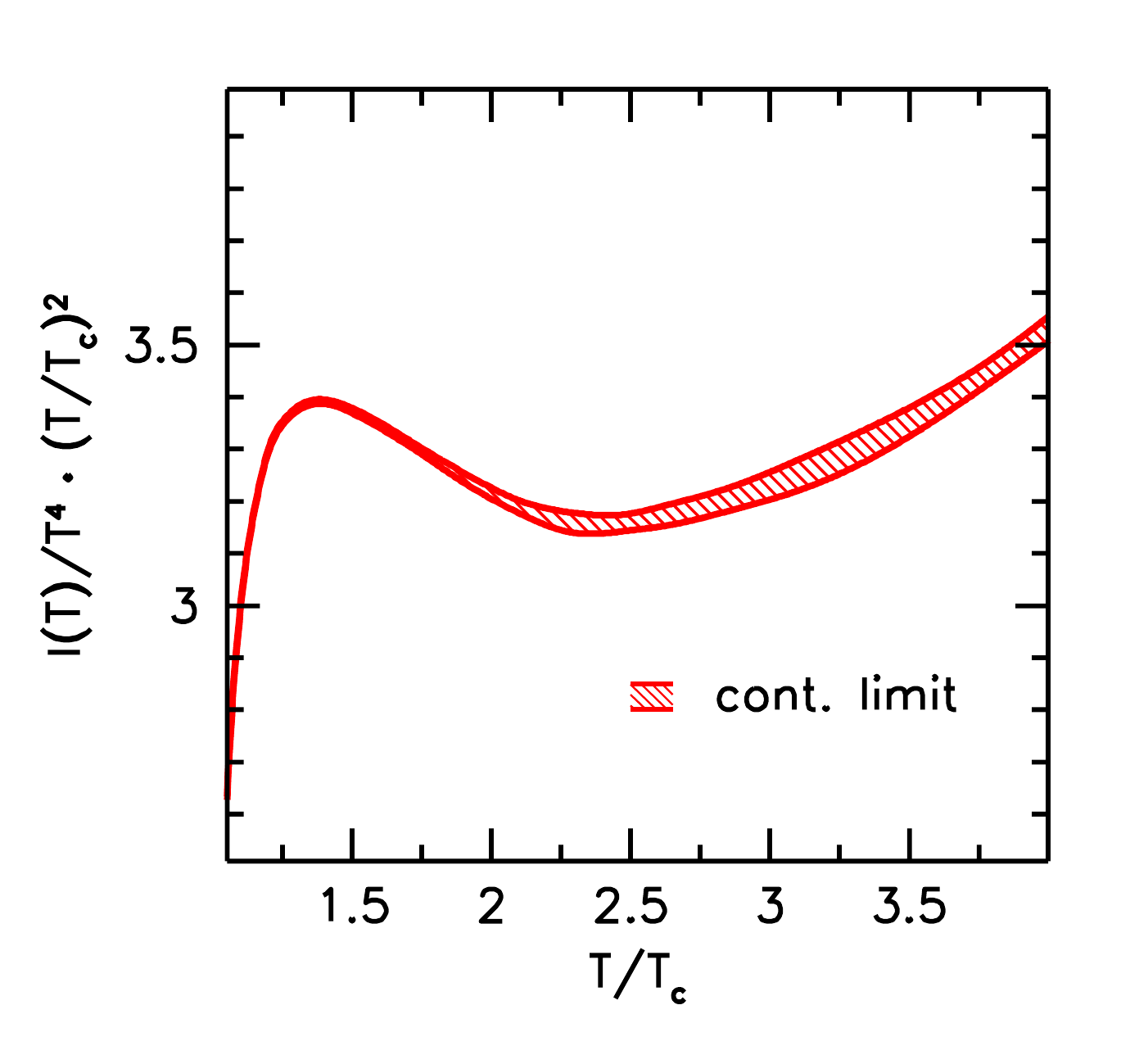}\hspace*{-0.3cm}
\includegraphics[width=11.13cm]{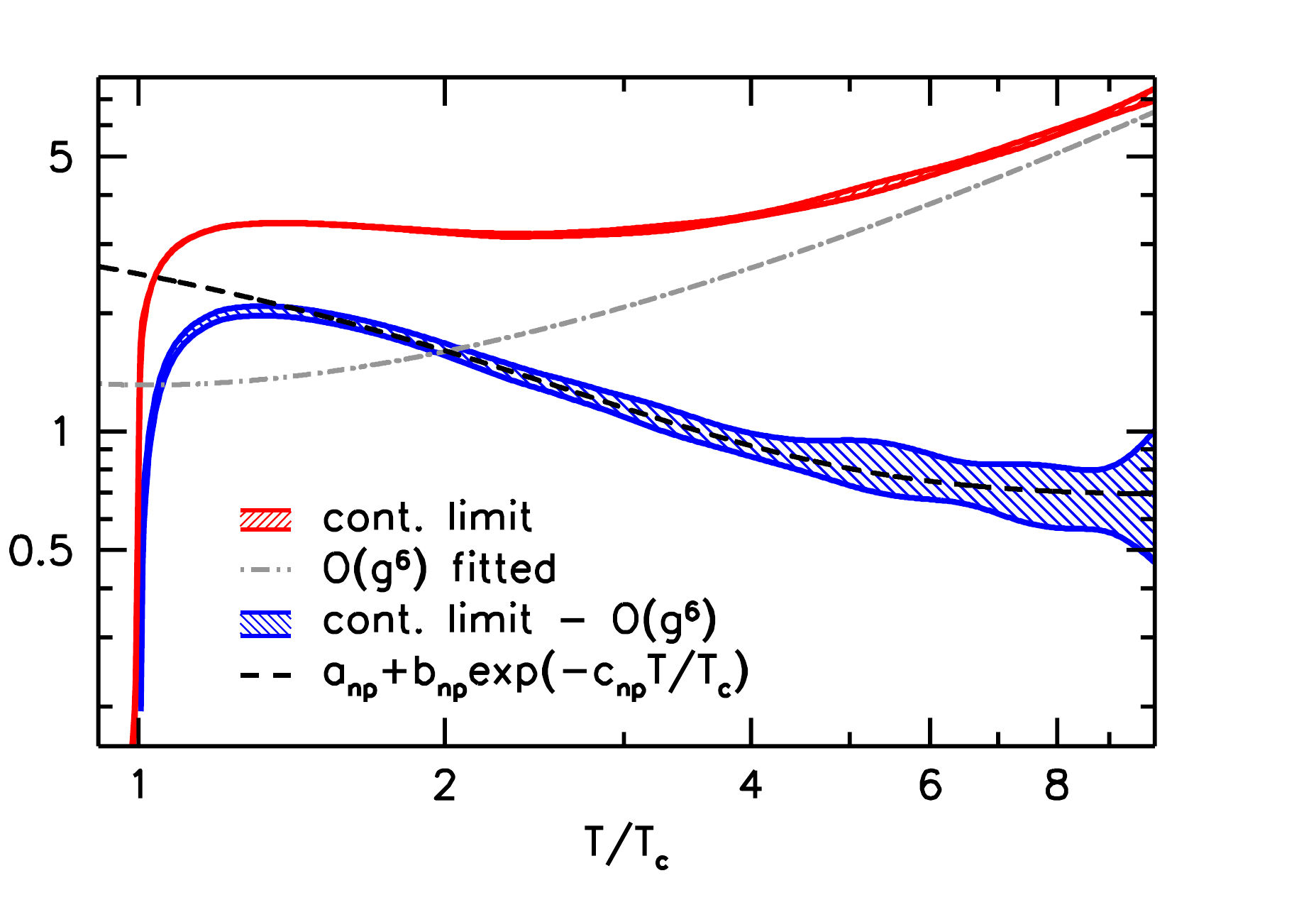}
}
\vspace*{-.8cm}
\caption{
Left panel: the continuum extrapolated lattice result for the $T^2$-scaled trace anomaly in the temperature region of $1.1\,T_c$ to $4\,T_c$. 
The data stays constant within a range of $5\%$ around the value $3.3$.
Right panel: the continuum limit obtained from the lattice results (red band), compared to fitted perturbation theory. We fit the $g^6$ coefficient (gray dashed-dotted line) and subtract it from the lattice results (blue band) to show the non-perturbative contribution which is then fitted by a simple function (black dashed line). }
\label{fig:pertcomp}
\end{figure}

Regardless of whether the conjecture of the last subsection is valid or not, we can make use of our small volume simulations at high temperature to compare to perturbative expansions, in particular, to extract some unknown coefficients of these formulas. We perform the continuum extrapolation in the same manner as for the large volume data, see section~\ref{sec:results}, using the $N_t=5,6$ and $8$ lattices. First we compare our results to $\mathcal{O}(g^6)$ improved perturbation theory~\cite{Kajantie:2002wa}. We perform a fit to the subtracted trace anomaly,
\be
\frac{I_{\rm pert}(T,q_c,\mu)}{T^4}-\frac{I_{\rm pert}(T/2,q_c,\mu)}{(T/2)^4},
\label{eq:subtrtracea}
\ee
for the unknown coefficient $q_c$ of the $g^6$ term with a fixed renormalization scale of $\mu=2\pi T$. If we also allow for a variation of the scale, we find $\mu/2\pi T$ to be consistent with 1 within errors. These fits are carried out for our results between $10\,T_c < T < 1000 \,T_c$ and the systematic error is estimated by varying the endpoints of the fit interval. Beyond this we also consider as a source of systematic error the uncertainty of our lattice scale setting (see section~\ref{sec:scaleset}). We quote as our final result for this parameter,
\be
q_c =-3526(4)(55)(30),
\label{eq:qcfit}
\ee
with the numbers in the parentheses are from left to right the statistical error, the error coming from the lattice scale and that from the variation of the fit interval. A good fit quality is indicated as $\chi^2/{\rm dof}=0.7$. The fitted function is shown by the dashed-dotted gray line in Fig.~\ref{fig:pertcomp}.
Note that a similar fit was attempted in the framework of an effective field theory in Ref.~\cite{Hietanen:2008tv}. Here we fitted the missing coefficient in the re-expanded formula (\ref{eq:ppert}), which we could, because our data stretches to temperatures where this re-expansion is justified.

While the $\sim T^{-2}$ behavior of the trace anomaly in the low-temperature region has been seen and studied in many papers (see e.g.~\cite{Boyd:1996bx,Meisinger:2001cq,Pisarski:2006yk,Andreev:2007zv,Panero:2009tv,Pisarski:long} and references therein), its relative weight in the total observable has not yet been quantified.
Therefore we also consider it useful to estimate the non-perturbative contribution to the trace anomaly, which we assume to be of the form $I_{\rm np}(T)/T^2T_c^2 = a_{\rm np} + b_{\rm np}\, e^{-c_{\rm np} T/T_c}$, i.e. we propose the following fit function:
\be
\frac{I_{\rm pert}(T)}{T^4} + \frac{a_{\rm np} + b_{\rm np}\exp(-c_{\rm np} \cdot T/T_c)}{(T/T_c)^2}.
\label{eq:fitNP}
\ee
First we perform the fit to our large volume $N_s/N_t=16$ continuum results for $a_{\rm np}$ with $b_{\rm np}=0$ kept fixed, then we carry out the fit for both non-perturbative coefficients. The fit interval is chosen to be $1.5\, T_c < T < 10\, T_c$. We find that the constant approximation is not able to resolve the trace anomaly in the low temperature region as $\chi^2/{\rm dof}\approx 25$. The exponential correction significantly improves the situation and we get $\chi^2/{\rm dof}=0.9$. Moreover, the parameters are rather sensitive to the variation of the lower endpoint of the fit interval which is just above the transition. Nevertheless, since there is no a priori constraint on the form of the fit function~(\ref{eq:fitNP}), we accept this as a first approximation to the non-perturbative contribution. We obtain the following coefficients:
\be
a_{\rm np} = 0.69(1)(9), \quad\quad b_{\rm np} = 3.64(3)(7),\quad\quad c_{\rm np} = 0.69(1)(2),
\ee
with the errors coming from the statistics and the lattice scale, respectively.
We also show this non-perturbative fit by the dashed black line in Fig.~\ref{fig:pertcomp}, in comparison with the lattice results minus the $\mathcal{O}(g^6)$ fitted formula.
Note that instead of using the second term of Eq.~(\ref{eq:fitNP}) an equally good description of our data can be given by a term of the form: $I_{\rm np}/T^4=A/T^2+B/T^3+C/T^4$.

\begin{figure}[h!]
\centering
\vspace*{-0.2cm}
\includegraphics[width=12.cm]{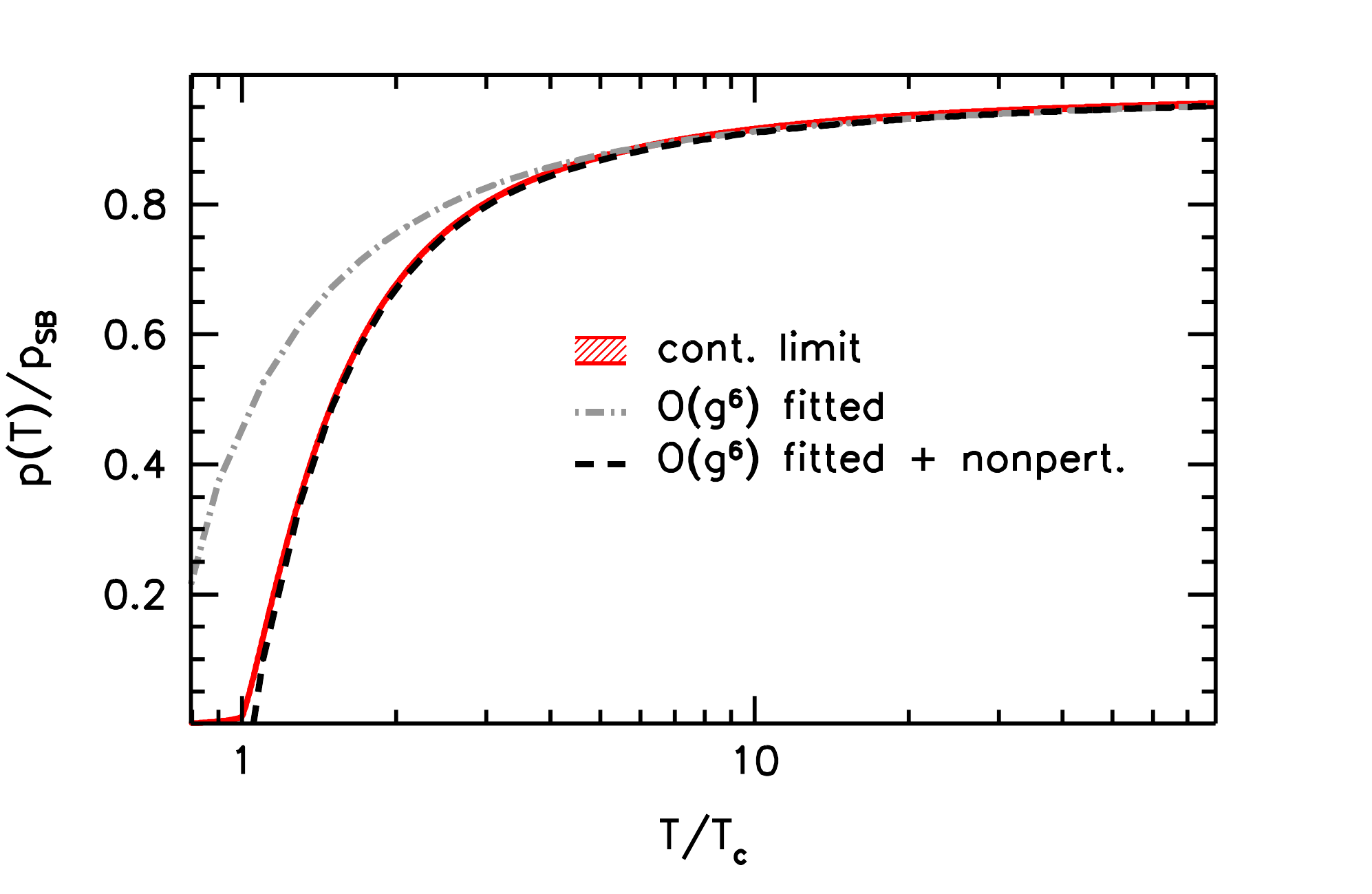}
\caption{The normalized pressure in the continuum limit. A comparison is shown to fitted $\mathcal{O}(g^6)$ perturbation theory and to perturbation theory plus an additional non-perturbative contribution (see text).}
\label{fig:pertcompp}
\end{figure}

\begin{figure}[h!]
\centering
\vspace*{-0.2cm}
\includegraphics[width=12.cm]{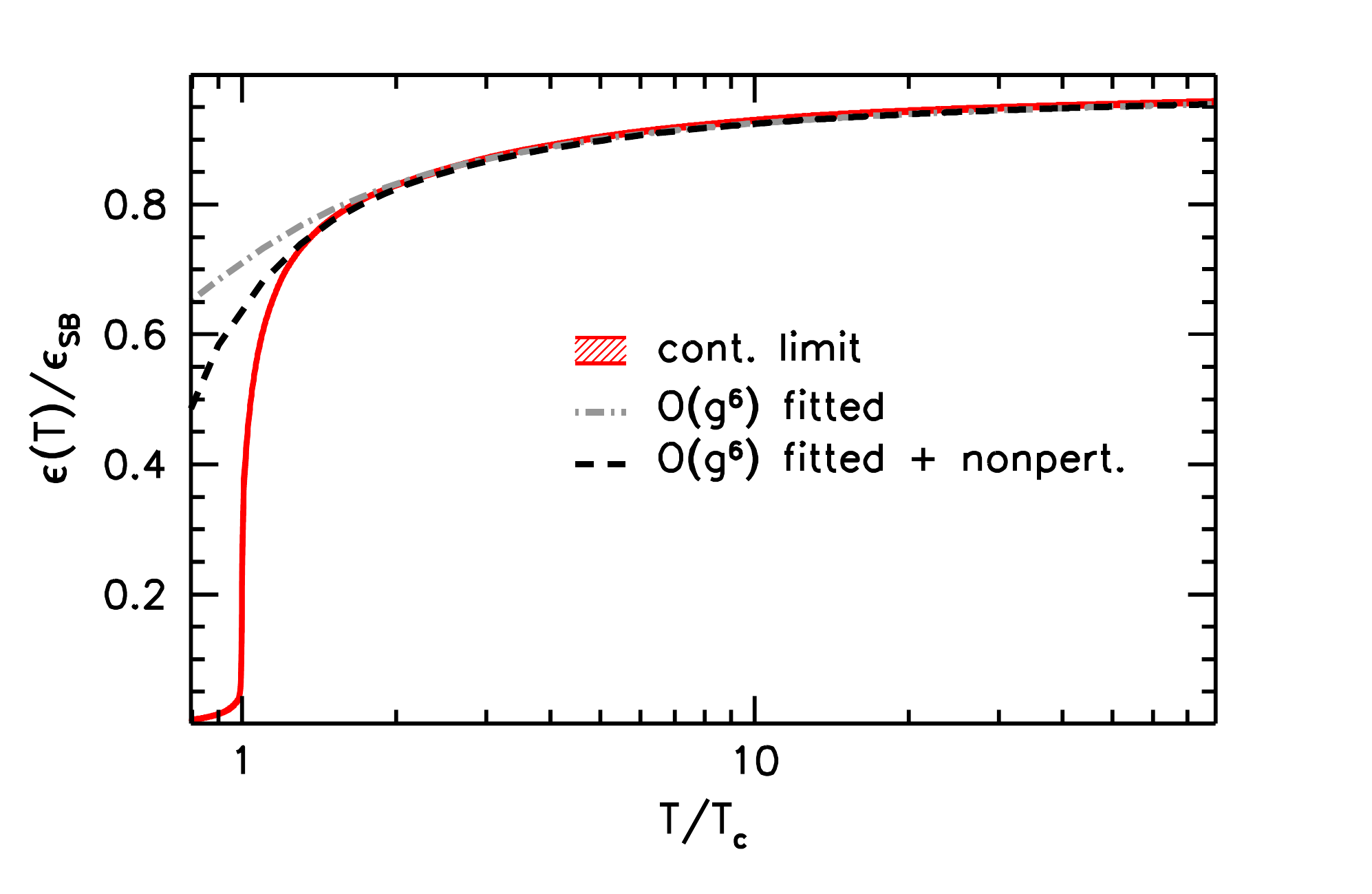}
\caption{The normalized energy density in the continuum limit. A comparison is shown to fitted $\mathcal{O}(g^6)$ perturbation theory and to perturbation theory plus an additional non-perturbative contribution (see text).}
\label{fig:pertcompe}
\end{figure}

\begin{figure}[h!]
\centering
\vspace*{-0.2cm}
\includegraphics[width=12.cm]{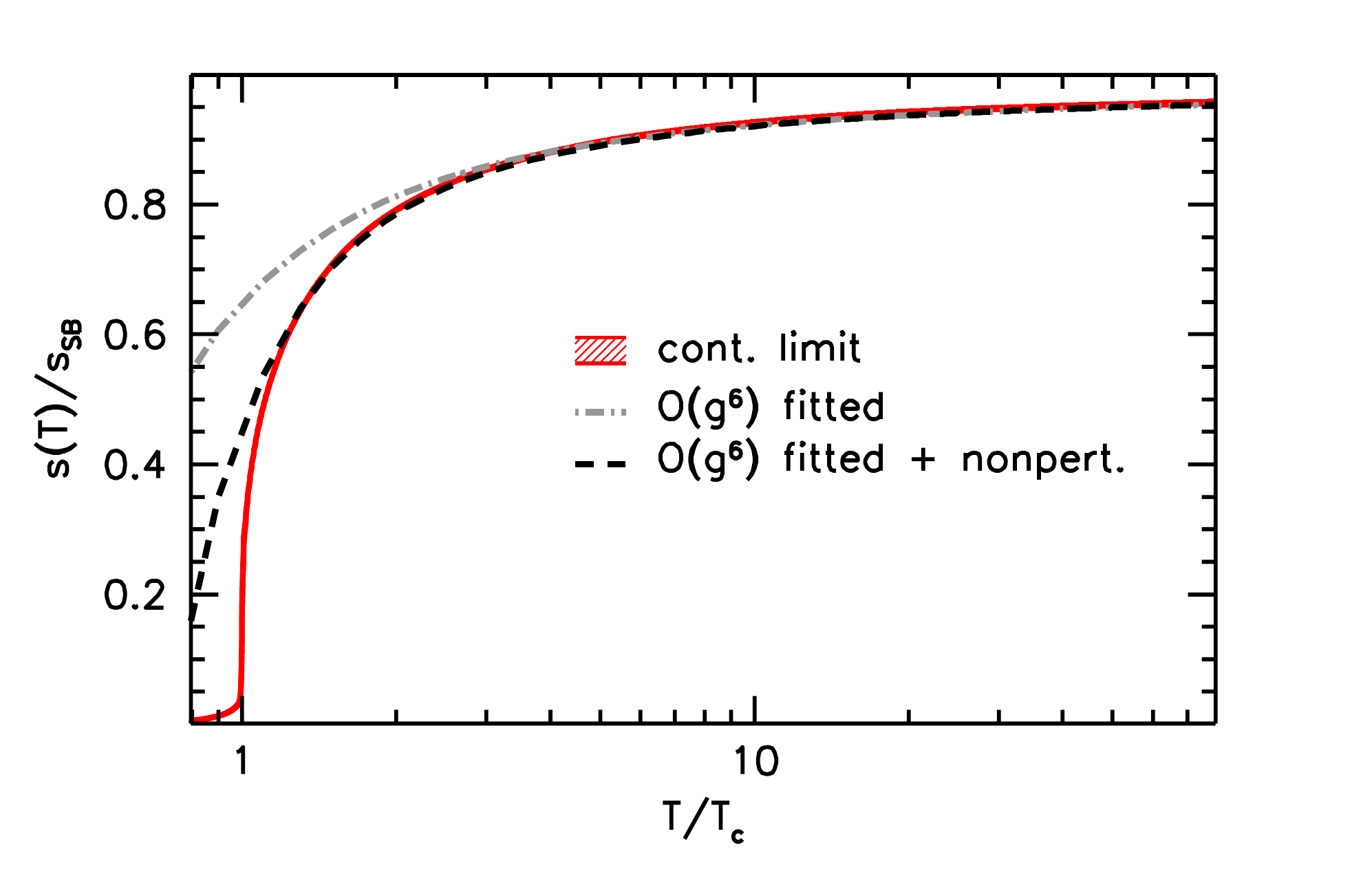}
\caption{The normalized entropy density in the continuum limit. A comparison is shown to fitted $\mathcal{O}(g^6)$ perturbation theory and to perturbation theory plus an additional non-perturbative contribution (see text).}
\label{fig:pertcomps}
\end{figure}

Using~(\ref{eq:pressureformula}) the fitted perturbative formulae for the pressure are also straightforward to write down. In Fig.~\ref{fig:pertcompp} we compare our continuum results to the so obtained predictions. Similar comparisons can be made for the case of the energy density and the entropy density also, where we find qualitatively the same behavior as for the pressure, see Figs.~\ref{fig:pertcompe} and~\ref{fig:pertcomps}. In these plots the large volume ($N_s/N_t=16$) continuum results are shown up to $10\,T_c$, continued with the small volume ($N_s/N_t=8$) continuum results beyond. The results for the trace anomaly and for the pressure are also listed up to $T/T_c=1000$ in table~\ref{tab1}.

\subsection{Fitting HTL perturbation theory}
\label{sec:fithtl}

\begin{figure}[ht!]
\centering
\vspace*{-0.2cm}
\includegraphics[width=12.cm]{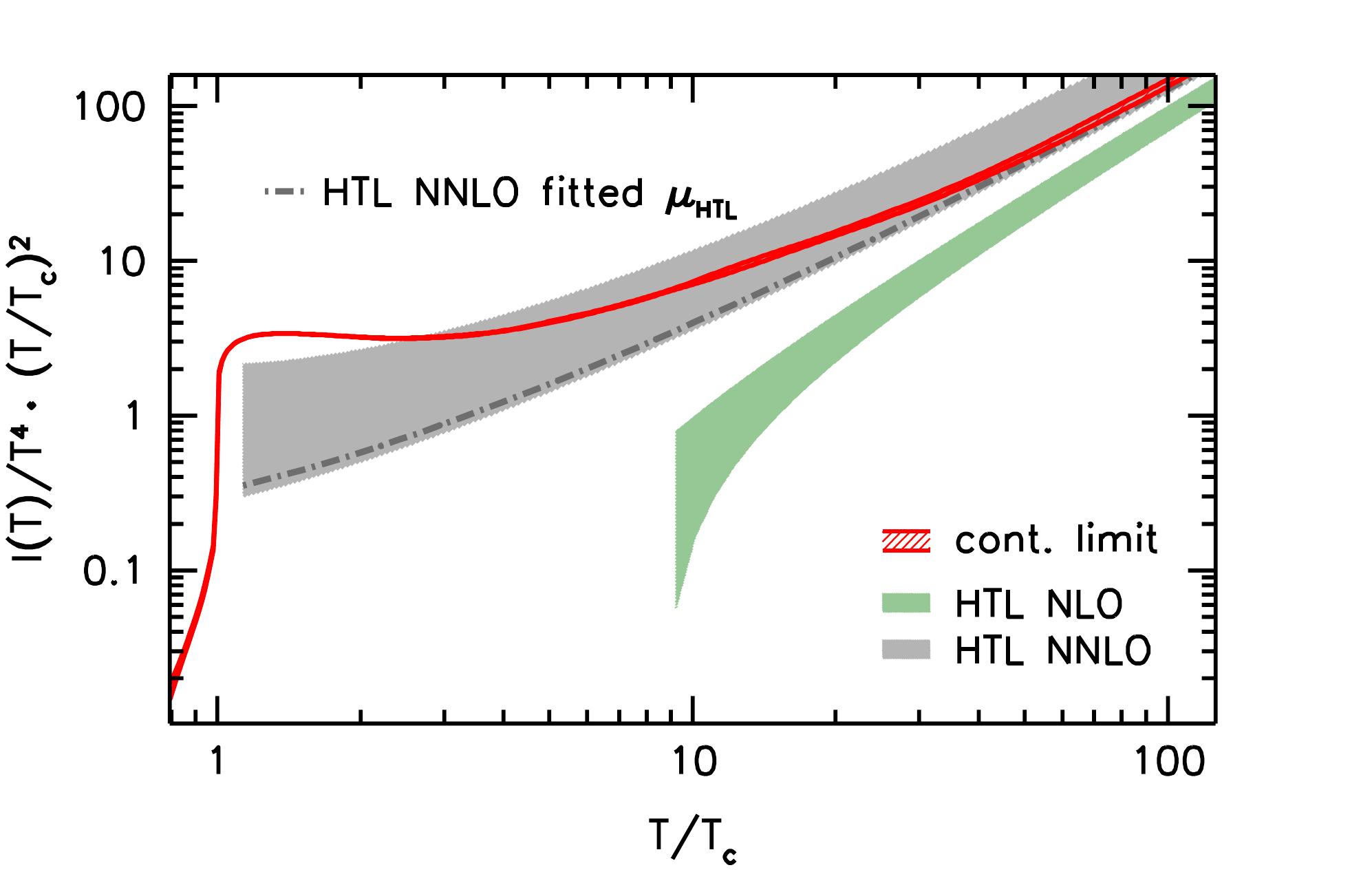}
\caption{The trace anomaly in the continuum limit, compared to the NLO and NNLO HTL expansion with varied renormalization scale $0.5<\mu_{\rm HTL}/2\pi T<2$ (green and gray shaded regions). The dashed-dotted line represents the NNLO expansion with the fitted scale (see text).}
\label{fig:htlcomp}
\end{figure}

Next we discuss the region of validity of the HTL resummed perturbation theory. In particular, we compare once again our $N_s/N_t=8$ continuum results to the NNLO expansion of the HTL scheme~\cite{Andersen:2010ct}. We consider the renormalization scale $\mu_{\rm HTL}$ as a free parameter of this expansion, and perform a fit to this parameter, i.e. our fit function to the subtracted trace anomaly is
\be
\frac{I_{\rm pert}(T,\mu_{\rm HTL})}{T^4}-\frac{I_{\rm pert}(T/2,\mu_{\rm HTL})}{(T/2)^4}.
\ee
The fit is carried out for $T>100T_c$, and the endpoint is varied to obtain the systematic error coming from the fitting procedure. The sum of deviations for this fit is $\chi^2/{\rm dof}=0.6$, indicating a nice agreement between lattice results and the perturbative expansion. Our result for the renormalization scale is (in the same notation for the errors as before)
\be
\frac{\mu_{\rm HTL}}{2\pi T} = 1.75(2)(6)(50).
\label{eq:fittedhtlmu}
\ee
The fitted formula for the trace anomaly is shown in Fig.~\ref{fig:htlcomp}.

\section{Theoretical description and model building}
\label{sec:models}

The present paper summarizes a long term project of us (for earlier reports see \cite{Endrodi:2007tq,Borsanyi:2011zm}). We have determined the equation of state of the pure SU(3) theory with a.) unprecedented accuracy and b.) in a far larger temperature range than previous studies. These two ingredients allow one to have a complete theoretical description of the equation of state from $T=0$ all the way to the phase transition, through the transition region into the perturbative regime up to the Stefan-Boltzmann limit. Our precision data will hopefully contribute to an even better understanding of the theory and/or model building. Below we summarize the various temperature regimes. These regimes can be described with different theoretical rigor and accuracy, which we comment on. First we discuss the confining phase, then the perturbative regime. The next region, which we study is the one above $T_c$ with its non-perturbative/non-ideal contribution. Finally comparing the latter with the confining phase we estimate the latent heat.

{\it i. Confining phase.} We provided a continuum extrapolated equation of state also in this phase. We found a nice agreement with Ref. \cite{Meyer:2009tq} (which is not continuum extrapolated yet, but the results are obtained on quite fine lattices), which also provided a Hagedorn-type description of its data up to the vicinity of $T_c$. It is observed that the gas model of stable gluons underestimates the equation of state below $T_c$ (c.f. Fig.~\ref{fig:coldtra}). Extending the spectral sum with an exponential spectrum $\rho(M)\propto \exp(M/T_h)$ (suggested by Hagedorn \cite{Hagedorn:1965st} almost half a century ago) provides a good description of the lattice result (see Fig. \ref{fig:hagedorn}). A simple fit to the Hagedorn model between the first and the last simulation points of Ref.~\cite{Meyer:2009tq}, i.e. between $0.7\,T_c$ and $0.985\,T_c$, can provide an accurate description of the equation of state on the few percent level,
\be
\frac{s_{\rm conf}(T)}{T^3}=
-0.2\cdot \frac{T}{T_c} - 0.134\cdot \log \left( 1.024 - \frac{T}{T_c}\right).
\label{eq:scoldparam}
\ee
Here we used $T_h/T_c=1.024(3)$ of Ref.~\cite{Meyer:2009tq} as a fixed parameter.
Our (preliminary) data at low temperatures has been put
into the context of various gauge algebras in Ref.~\cite{Buisseret:2011fq}.

{\it ii. Perturbative regime and the Stefan Boltzmann limit.}
We have determined the unknown coefficient of the $g^6$ term of the perturbative approach. The perturbative result with this $g^6$ term is accurate already from about $10\,T_c$ all the way up to the Stefan-Boltzmann limit (see our discussion in Sec.~\ref{sec:fitg6}). The equation of state in this expansion contains various terms of log($T/T_c$) and $g$ (which can be expressed by logarithms of $T/\Lambda_{\overline MS}$ or $T/T_c$, too). As a quick reference we provide the normalized pressure $p(T)/T^4$ as a function of the strong coupling $\alpha_s$ to order $\alpha_s^3\log \alpha_s$~\cite{Kajantie:2002wa},
\be
\begin{split}
\frac{p_{\rm pert}}{T^4} = \frac{8\pi^2}{45} \Big[
&1
-1.1937\cdot\alpha_s
+5.3876\cdot\alpha_s^{3/2}
+16.2044\cdot\alpha_s^{2}
+6.8392\cdot\alpha_s^{2}\cdot\log(\alpha_s) \\
&-45.6800\cdot\alpha_s^{5/2}
-36.5990\cdot\alpha_s^{3}\cdot\log(\alpha_s)
+41.8960\cdot\alpha_s^{3}
+0.03225\cdot q_c\cdot\alpha_s^3
\Big],
\end{split}
\label{eq:ppert}
\ee
where in the last term the result of our fit to $q_c$, Eq.~(\ref{eq:qcfit}) enters,
$q_c=-3526(4)(55)(30)$.
For the coupling constant one may use the three-loop formula~\cite{Nakamura:2010zzi}, at renormalization scale $\mu=2\pi T$,
\be
\alpha_s(T) = 
1.1424 \cdot \frac{1}{t}
-0.9630 \cdot\frac{\log t}{t^2}
+0.4143 \cdot\frac{1}{t^3}
-0.8118 \cdot\frac{\log t}{t^3}
+0.8118 \cdot\frac{(\log t)^2}{t^3},
\ee
with
\be
t=4.1380+2\cdot\log(T/T_c),
\ee
where we used the central value of our
$T_c/\Lambda_{\overline{MS}}=1.26(7)$ determination.
Clearly, from $p$ one can obtain all other thermodynamic observables.

{\it iii. Deconfined phase with non-perturbative contribution.}
As it was observed in \cite{Pisarski:2006hz,Pisarski:2006yk} the lattice data for the $T^2$ scaled trace anomaly is essentially constant in the temperature range $T\approx1.3-4\,T_c$. The author suggested an effective Lagrangian based on the Wilson-line ($L$), in which the confinement-deconfinement transition arises through the term $\propto T^2B_f\vert\textrm{tr}~L\vert^2$. Adding such a mass term is standard in Landau-Ginzburg type of analyses.
One needs a linear term, too (see e.g. Refs.~\cite{Dumitru:2010mj,Pisarski:long}), which was first suggested in Ref.~\cite{Meisinger:2001cq}).

The new data confirmed the existence of such a non-perturbative or non-ideal term, proportional to $T^2$. Subtracting the perturbative result from the lattice data one can determine this non-perturbative/non-ideal contribution. Interestingly enough this term has an exponentially decaying part. For completeness, we repeat the formula for this non-perturbative term here again:
\be
\frac{I_{\rm np}}{T^4}=\frac{T_c^2}{T^2}\left[ a_{\rm np}+b_{\rm np}\exp(-c_{\rm np}\cdot T/T_c)\right],
\ee
with $a_{\rm np}=0.69(1)(9)$, $b_{\rm np}=3.64(3)(7)$ and 
$c_{\rm np}=0.69(1)(2)$.
For this term we have chosen a form, in which the coincidence between the numerical values for $a_{np}$ and $c_{np}$ is transparent (this coincidence might be interesting from the model building point of view). The sum of the terms $I_{\rm pert}/T^4+I_{\rm np}/T^4$ describes the data down to about $1.3\,T_c$.

{\it iv. Phase transition.} The pure $\mathrm{SU}(3)$ gauge theory undergoes a weak first order phase transition. The strength of the phase transition is well illustrated by the dimensionless latent heat. Its value $L_h/T_c^4\approx1.4$ is fairly well known from the literature~\cite{Lucini:2005vg,Beinlich:1996xg,Meyer:2009tq}.

On the confining side of the transition the hadron resonance gas provides a good description up to the vicinity of $T_c$. This confining phase within the Hagedorn model (see the point {\it i.} of our discussion) ends with an entropy value of $s/T_c^3\approx0.3$ (note that the normalized pressure is much smaller).

As we have seen in the previous two points the perturbative approach with an intrinsically non-perturbative part describes the data from the Stefan-Boltzmann limit all the way down to the vicinity of the phase transition, to about $1.3\,T_c$. In this deconfined phase one observes an approximately constant behavior of the $T^2$ scaled trace anomaly, with a value around 3.3, see left panel of Fig.~\ref{fig:pertcomp}. One could naively extend this plateau to $T_c$ and take the appropriate difference between the energy densities of the two sides of the transition. 
Using this naive procedure one ends up with a latent heat, which is about twice as large as the real value, measured on the lattice. 
The reason for that is that the plateau in the trace anomaly turns down as it gets closer to $T_c$. On the other side of $T_c$ a similar (upward) effect appears.
Though a factor of two might seem large, with the new precision data in hand one hopes to understand more about the region between $T_c$ and $1.3\,T_c$ and to come up with even better models and results (see e.g. Ref.~\cite{Pisarski:long} and references therein). To that end the lattice should provide better data for the temperature dependence of the renormalized Polyakov loops, which will be the topic of a forthcoming publication.

\begin{acknowledgments}
This research has been partly supported by the Research Executive Agency (REA) of the European Union under Grant Agreement number PITN-GA-2009-238353 (ITN STRONGnet) as well as by DFG FO-502 and SFB-TR/55.
The simulations have mainly been performed on the QPACE facility. Part of the calculation was running on the
GPU~\cite{Egri:2006zm} clusters at the Wuppertal University and the E\"otv\"os University with the support from the
European Research Council grant 208740 (FP7/2007-2013). The authors acknowledge
the helpful comments from Axel Maas, Aleksi Kurkela, Marco Panero, Rob Pisarski, Kari
Rummukainen, York Schr\"oder and Mike Strickland.
\end{acknowledgments}

\bibliographystyle{JHEP}
\bibliography{su3_jhep}

\begin{center}
\mbox{
\begin{tabular}{|c|c|c||c|c|c|}
\hline
$T/T_c$ & $I/T^4$ & $p/T^4$ & $T/T_c$ & $I/T^4$ & $p/T^4$ \\
\hline\hline
0.70 & 0.0104(25) & 0.0015(1) & 	3.0 & 0.3589(27) & 1.4098(13) \\
0.74 & 0.0162(27) & 0.0023(0) & 	3.5 & 0.2736(20) & 1.4582(14) \\
0.78 & 0.0232(31) & 0.0033(1) & 	4.0 & 0.2207(11) & 1.4910(14) \\
0.82 & 0.0318(22) & 0.0046(2) & 	4.5 & 0.1855(15) & 1.5149(14) \\
0.86 & 0.0433(19) & 0.0064(3) & 	5.0 & 0.1606(21) & 1.5330(14) \\
0.90 & 0.0594(22) & 0.0087(3) & 	6.0 & 0.1266(13) & 1.5591(17) \\
0.94 & 0.0859(36) & 0.0118(3) & 	7.0 & 0.1050(10) & 1.5768(18) \\
0.98 & 0.1433(47) & 0.0164(4) & 	8.0 & 0.0903(9) & 1.5898(18) \\
1.00 & 1.0008(672) & 0.0222(4) & 	9.0 & 0.0798(8) & 1.5998(19) \\
1.02 & 2.0780(137) & 0.0571(9) & 	10 & 0.0720(15) & 1.6078(19) \\
1.06 & 2.4309(29) & 0.1455(10) & 	20 & 0.0375(16) & 1.6444(29) \\
1.10 & 2.4837(38) & 0.2370(10) & 	30 & 0.0265(13) & 1.6572(35) \\
1.14 & 2.4309(22) & 0.3250(10) & 	40 & 0.0216(11) & 1.6641(40) \\
1.18 & 2.3426(17) & 0.4074(10) & 	50 & 0.0191(11) & 1.6686(43) \\
1.22 & 2.2342(28) & 0.4837(10) & 	60 & 0.0174(12) & 1.6720(46) \\
1.26 & 2.1145(20) & 0.5539(10) & 	80 & 0.0154(12) & 1.6767(48) \\
1.30 & 1.9980(21) & 0.6181(9) & 	100 & 0.0142(12) & 1.6800(50) \\
1.34 & 1.8867(21) & 0.6770(9) & 	200 & 0.0112(11) & 1.6887(53) \\
1.38 & 1.7809(19) & 0.7309(9) & 	300 & 0.0100(12) & 1.6930(53) \\
1.42 & 1.6810(17) & 0.7804(9) & 	400 & 0.0091(12) & 1.6958(53) \\
1.46 & 1.5872(17) & 0.8258(9) & 	500 & 0.0085(12) & 1.6977(52) \\
1.5 & 1.4995(19) & 0.8675(9) & 	600 & 0.0080(12) & 1.6992(52) \\
2.0 & 0.8038(24) & 1.1890(8) & 	800 & 0.0073(11) & 1.7014(52) \\
2.5 & 0.5057(23) & 1.3319(12) & 	1000 & 0.0068(10) & 1.7030(52) \\
\hline
\end{tabular}

}
\begin{table}[ht!]
\caption{Continuum extrapolated lattice results for the trace anomaly and the pressure as functions of the temperature.
\label{tab1}
}
\end{table}
\end{center}

\end{document}